\documentclass{elsart}
\begin{document}

\begin{frontmatter}
\title{The Doppler shift in a Schwarzschild spacetime}

\author[PWr]{A. Radosz\corauthref{cor1}}
\ead{Andrzej.Radosz@pwr.wroc.pl}
\corauth[cor1]{Corresponding author}
\author[KU]{A.T. Augousti}
\ead{augousti@kingston.ac.uk}
\author[PWr]{K. Ostasiewicz}
\ead{Katarzyna.Ostasiewicz@pwr.wroc.pl}

\address[PWr]{Institute of Physics, Wroc{\l}aw University of Technology,
  Wybrze\.ze~Wyspia\'nskiego~27, 50-370~Wroc{\l}aw, Poland}
\address[KU]{Faculty of Science, Kingston University,
  Kingston, Surrey KT1 2EE UK}

\begin{abstract}
We demonstrate that in the case of Schwarzschild spacetime the Doppler shift
is partially factorized into terms representing relativistic, kinematical and
the gravitational contributions. The condition for the complete factorization
is derived. Application of these results to the simplest cases and possible
implementation in the framework of GPS is briefly discussed.
\end{abstract}
\end{frontmatter}

Light emitted by a moving source in a gravitational field is shifted in
frequency. This frequency shift is a combined effect consisting of
contributions arising from kinematical and gravitational origins. In general,
it is expressed in the form of entangled velocity components (see
\cite{Bahder,LT}). Consider a source A and a receiver B, traveling along an
arbitrary trajectory in a gravitational field; in this case the frequency
shift can be expressed in terms of the components of the relative velocity
between A and B. Explicitly, this takes the form,
\begin{equation}
  {\omega_{\rm B}\over\omega_{\rm A}}=
  {1\over\left(\sqrt{1-v^2_{\rm AB}}\right)^{-1}+v_R},
\label{gwiazdka}
\end{equation}
where quantities relating to the relative motion of A and B have been
introduced, namely the {\it speed\/} $v_{\rm AB}$ and the {\it recession
speed\/} $v_R$ \cite{Bahder} (see below).

One can see from Eq. (\ref{gwiazdka}) that different contributions to the
frequency shift have been entangled in such a way that the effect of the
gravitational field is not explicitly manifested (though see \cite{Bahder,LT}).
On the other hand, one might expect that under particular symmetries (for
instance, of the spacetime geometry or of particular trajectories within a
given spacetime geometry) a decoupling of (at least) some of the effects
contributing to the total frequency shift may occur. Such a decoupling is
expected to be revealed in a factorized form of the total Doppler shift
(\ref{gwiazdka}).

In this paper we confine ourselves to a discussion of the case of a
Schwarzschild spacetime geometry. In such a case one can identify three
different forms of frequency shift, namely gravitational (blue/red) shift and
the shift arising from kinematical effects (classical and relativistic). The
expression for the frequency shift is derived here in the case of arbitrary
motion of observers A and B exchanging light signals and it is expressed in
terms of these three contributions. It is found that due to the spacetime
symmetry a partial decoupling occurs, namely via factorization of the
relativistic contributions, the kinematical one and the gravitational shift.
The kinematical factor, time-dilation-like, appears in inverse form. The
velocity components of observers A and B and the wave vector components for
the light signal are mixed in the form of scalar product, leading to a
particular type of coupling. It is suggested that this result might be of
interest in the context of Satellite Navigational Systems
\cite{Bahder,LT,GPS1,GPS2}. It is shown that {\it complete factorization\/} into
gravitational and kinematical contributions occurs for a special class of
trajectories for A and B --- in this case the conventional form of time
dilation factor may be restored. A similar factorization also pertains for a
particular in-falling trajectory in the case of a Kerr geometry.

Let us begin the discussion from the Schwarzschild spacetime geometry, where
the line element expressed in Schwarzschild coordinates takes the form:
\begin{eqnarray}
  ds^2&=&\left(1-\frac{r_S}{r}\right)dt^2-
  \left(1-{r_S\over r}\right)^{-1}dr^2-r^2d\theta^2-
  r^2\sin^2\theta\,d\varphi^2
\nonumber
\\
  &\equiv& g_{tt}dt^2+g_{rr}dr^2+g_{\theta\theta}d\theta^2+
  g_{\varphi\varphi}d\varphi^2.
\label{schw}
\end{eqnarray}
There are two Killing vectors in this case, $\tilde\eta$ and $\tilde\xi$,
\begin{equation}
  \eta^\alpha=\delta^\alpha_t, \quad \xi^\alpha=\delta^\alpha_\varphi,
\label{killing}
\end{equation}
corresponding to energy and angular momentum conservation. A static observer S
provides the absolute rest system, and his velocity is the normalized Killing
vector $\tilde\eta$ (see, e.g., \cite{abramowicz}):
\begin{equation}
  \tilde N=(\eta^\alpha\eta_\alpha)^{-1/2}\tilde\eta=
  (g_{tt}^{-1/2})\tilde\eta.
\label{en}
\end{equation}
A {\it massless\/} particle (e.g., a photon) follows a (planar) null geodesic,
\begin{equation}
  k^\alpha k_\alpha =0,
\label{ka}
\end{equation}
where the tangential vector, the {\it wave vector}, is denoted as $\tilde k$.
Its time component is conserved,
\begin{equation}
  k_t=\omega_\infty,
\label{kat}
\end{equation}
where $\omega_\infty$ is the frequency observed at infinity. An observer B
follows a world line, characterized by a time-like velocity vector
$\tilde U$:
\begin{equation}
  U^\alpha U_\alpha=1.
\label{uu}
\end{equation}
Receiving a light signal at $\tilde{x}_{\rm B}$, B measures its frequency as
the time-component of the $\tilde k$ vector:
\begin{equation}
  \omega_{\rm B}=U^\alpha k_\alpha=U^t k_t\left(1+{U^ik_i\over U^tk_t}\right),
\label{omb}
\end{equation}
where the notation,
\begin{equation}
  U^ik_i\equiv g_{rr}U^rk^r+g_{\varphi\varphi}U^\varphi k^\varphi
\label{omb2}
\end{equation}
has been applied (geodesic (\ref{ka}) might be confined to an equatorial
plane). The frequency shift is given as the ratio of the frequency of the
light signal measured by B, Eq. (\ref{omb}), compared to its emission
frequency. In order to express this in terms of gravitational and kinematical
shifts, let us first consider the case of a distant, static source for the
light signal, then the emission frequency is $\omega_\infty$. The velocity
$\tilde V$ of observer B with respect to the static observer S, is given by
(see, e.g., \cite{abramowicz}):
\begin{equation}
  \tilde V=\tilde U-(U^\alpha N_\alpha)\tilde N=(0,U^i)
\label{e10a}
\end{equation}
and is related to the {\it speed}, $v_{\rm B}$, (as measured by S) as:
\begin{equation}
  V^\alpha V_\alpha=-{v^2_{\rm B}\over 1-v^2_{\rm B}}\equiv
  -(\gamma_{\rm B}v_{\rm B})^2,
\label{e10b}
\end{equation}
where $\gamma_{\rm B}$ is the usual time-dilation factor. Using (\ref{e10a})
one finds that the time component $U^t$, may be expressed in the form of a
product of gravitational shift and time dilation factors (see also
\cite{APPB}),
\begin{equation}
  U^t={1\over\sqrt{g_{tt}(r_{\rm B})}}\gamma_{\rm B}.
\label{e11}
\end{equation}
Further simplification of expression (\ref{omb}) results from the fact that
the ratio within the bracket in Eq.~(\ref{omb}) takes the form of the
(pseudo)scalar product. Indeed, introducing the following {\it
three-vectors\/} $\vec v$ and $\hat{\kappa}$:
\begin{equation}
  v^i\equiv {U^i\over \sqrt{g_{tt}}U^t},
\label{e12a}
\end{equation}
\begin{equation}
  \kappa^i\equiv {k^i \over \sqrt{g_{tt}}k^t},
\label{e12b}
\end{equation}
and applying (\ref{uu}) and (\ref{e10a}),
\begin{equation}
  v^2=-{U^iU_i\over U^tU_t}
\label{e12c}
\end{equation}
one can express the ratio in Eq. (\ref{omb})
\begin{equation}
  {U^ik_i\over U^tk_t}=v^i\kappa_i
\label{e13}
\end{equation}
as a (pseudo) scalar product of the {\it velocity}, $\vec{v}$, and {\it
tangential\/} ({\it unit\/}), $\hat{\kappa}$, three-vectors. Therefore, the
frequency shift (\ref{omb}), with respect to a static, distant source, takes
the partially decoupled form:
\begin{equation}
  {\omega_{\rm B}\over\omega_\infty}=
  {1\over\sqrt{g_{tt}(r_{\rm B})}}\gamma_{\rm B}(1+v_{\rm B}^i\kappa_i).
\label{e14}
\end{equation}
Generalization to the case of an arbitrary emission event A is
straightforward. Source A follows a world line, characterized by a velocity
vector $\tilde U_{\rm A}$, and at position $\tilde{x}_{\rm A}$ emits a light
signal characterized by wave vector $k'$. Then the emission frequency is given
by (see (\ref{omb}), (\ref{e14}))
\begin{equation}
  \omega_{\rm A}=
  \omega'_\infty{1\over\sqrt{g_{tt}(r_{\rm A})}}
  \gamma_{\rm A}(1+v^i_{\rm A}\kappa'_i).
\label{e15}
\end{equation}
As B receives a signal emitted by A,
\[
  \omega_\infty=\omega'_\infty
\]
and the frequency shift, defined as a ratio of reception frequency (reception
event by B) and emission frequency (emission event by A), is finally given as
\begin{equation}
  {\omega_{\rm B}\over\omega_{\rm A}}=
  {\sqrt{g_{tt}(r_{\rm A})}\over\sqrt{g_{tt}(r_{\rm B})}}
  {\gamma_{\rm B}\over\gamma_{\rm A}}
  {(1+v_{\rm B}^i\kappa_i)\over (1+v_{\rm A}^i\kappa'_i)}.
\label{e16}
\end{equation}
As noted above, partial factorization of this expression into gravitational,
$1/\sqrt{g_{tt}(r)}$, and kinematical, $\gamma$, contributions is the result
of the symmetry of this spacetime (see also \cite{APPB}). It should
also be noted that the time dilation factor that appears in (\ref{e16}) is in
an inverse form. The last factor on the right hand side (\ref{e16}),
expresses auxiliary coupling of A and B velocity components via scalar
products with tangential (unit) wave vectors.

The frequency shift, as given in (\ref{e16}), is applicable in the case of
Schwarzschild spacetime, and here we would like to emphasize its manifestation
in some simplest cases. In particular situations it may be further simplified.
This occurs when the velocity vectors, $\vec v$ and a tangential unit wave
vector $\hat\kappa$ are {\it parallel\/} (or anti-parallel), or {\it
orthogonal}.

In the first case
\begin{equation}
  (v^i\kappa_i)^2=(v^iv_i)\cdot (\kappa^i\kappa_i)=v^2
\end{equation}
and
\begin{equation}
  v^i\kappa_i=\mp v,
\label{e17}
\end{equation}
where ``$-$'' and ``$+$'' correspond to the parallel and anti-parallel
orientation of $\vec v$ and $\hat\kappa$, respectively.

In the case of A and B travelling tangentially to the light ray during
emission and reception of the signal, respectively, $\vec v$ and
$\hat\kappa$ are {\it parallel}, and (\ref{e16}) simplifies to:
\begin{equation}
  {\omega_{\rm B}\over\omega_{\rm A}}=
  {\sqrt{g_{tt}(r_{\rm A})}\over\sqrt{g_{tt}(r_{\rm B})}}
  {\gamma_{\rm A}\over\gamma_{\rm B}}{(1+v_{\rm A})\over(1+v_{\rm B})}.
\label{e18}
\end{equation}
In this case the frequency shift is expressed in a factorized form of the
ratios of relativistic contributions --- gravitational shifts and time
dilations --- and classical Doppler shifts (the last term on the right hand
side of (\ref{e18})); the conventional form of the time dilation is restored.

In the second case, of orthogonal vectors $\vec v$ and $\hat\kappa$,
\begin{equation}
  v^i\kappa_i=0.
\label{e19}
\end{equation}
The frequency shift factorizes into gravitational shift and inverse
time-dilation contributions:
\begin{equation}
  {\omega_{\rm B}\over\omega_{\rm A}}=
  {\sqrt{g_{tt}(r_{\rm A})}\over\sqrt{g_{tt}(r_{\rm B})}}
  {\gamma_{\rm B}\over\gamma_{\rm A}}.
\label{e20}
\end{equation}
It is worth identifying two special applications of (\ref{e18}).

In the case of equal velocities of the source, $v_{\rm A}$, and detector,
$v_{\rm B}$, the frequency shift simplifies solely to the gravitational shift,
\begin{equation}
  {\omega_{\rm B}\over\omega_{\rm A}}=
  {\sqrt{g_{tt}(r_{\rm A})}\over\sqrt{g_{tt}(r_{\rm B})}}.
\label{e21}
\end{equation}
Both velocities $v_{\rm A}$ and $v_{\rm B}$ are measured with respect to the
local static observers arranged at $r_{\rm A}$ and $r_{\rm B}$, respectively.

In the case of a radial fall, an in-falling observer B, who starts free fall
at $r_{\rm A}$ and receives a light signal emitted from the starting point A, finds
a classical-like frequency shift (see also \cite{AR}):
\begin{equation}
  {\omega_{\rm B}\over\omega_{\rm A}}={1\over 1+v_{\rm B}}.
\label{e22a}
\end{equation}
On the other hand, signals sent by B towards observer A, are received as
red-shifted (see (\ref{e18})):
\begin{equation}
  {\omega_{\rm A}\over\omega_{\rm B}}=1-v_{\rm B}.
\label{e22b}
\end{equation}
When B approaches the critical radius $r_S$, its velocity goes to 1 as
$v_{\rm B}\propto\sqrt{r_S/r}$ and redshift (\ref{e22a}) and redshift
(\ref{e22b}) are expressed as $\left(1+\sqrt{r_S/r}\right)^{-1}$ and
$1-\sqrt{r_S/r}$, respectively.

The status of result (\ref{e16}) is as follows. It has been derived for a
particular geometry, therefore it should be regarded as a particular case of a
more general formula, (\ref{gwiazdka}). The verification, though
straightforward, is rather lengthy and we illustrate it in the simplest case
of radial free fall (\ref{e22a}). In expression (\ref{gwiazdka}) key feature
is A's four-velocity,
\begin{equation}
  \tilde U_{\rm A}(r_{\rm A})={1\over\sqrt{g_{tt}(r_{\rm A})}}\delta^\alpha_t
\end{equation}
parallel transported to $r_{\rm B}$:
\begin{equation}
  U^t_{\rm A}(r_{\rm B})={1\over 2}(a+b)
  \qquad
  U^r_{\rm A}(r_{\rm B})={1\over 2}g_{tt}(r_{\rm B})(a-b)
\end{equation}
\begin{equation}
  a={\sqrt{g_{tt}(r_{\rm A})}\over g_{tt}(r_{\rm B})}
  \qquad
  b={1\over\sqrt{g_{tt}(r_{\rm A})}}.
\end{equation}
Its scalar product with $\tilde U_{\rm B}$ and with a unit vector, $\tilde R$
orthogonal to $\tilde U_{\rm B}$,
\begin{equation}
  \tilde R(r_{\rm B})=
  {1\over g_{tt}(r_{\rm B})}(U^r_{\rm B},g_{tt}(r_{\rm B})U^t_{\rm B})
\end{equation}
yields a {\it relative speed\/} term, $1/\sqrt{1-v_{\rm AB}^2}$, and a
so-called {\it recession speed\/} $v_R$ (see \cite{Bahder}), respectively.

Then inserting both these scalar products into (\ref{gwiazdka}) one obtains
\begin{equation}
  {1\over \left(\sqrt{1-v_{\rm AB}^2}\right)^{-1}+v_R}
  ={1\over
    1+\sqrt{{g_{tt}(r_{\rm A})-g_{tt}(r_{\rm B})\over g_{tt}(r_{\rm A})}}},
\end{equation}
i.e., the result (\ref{e22a}).

Let us discuss the broader context of the main result (\ref{e16}).
The approach presented above might be of wider practical use. Indeed,
following Bahder's \cite{Bahder} treatment of ``relativity of GPS'' (see also
\cite{GPS1,GPS2}), the metric applied in that case (Earth-centered inertial)
is:
\begin{eqnarray}
  ds^2&=&\left(1+{2\over c^2}(V-\varphi_{\rm O})\right)(dx^0)
\nonumber
\\
  &&{}-\left(1-{2\over c^2}V\right)^{-1}
    \left((dx^1)^2+(dx^2)^2+(dx^3)^2\right),
\label{dwie}
\end{eqnarray}
where $V$ and $\varphi_{\rm O}$ denote Earth's potential and geopotential on
the equator. This metric is time independent and diagonal. In such a case our
method and result (\ref{e16}) might be implemented --- the difference between
the approach presented here and those discussed in \cite{Bahder,GPS1,GPS2} is
obvious: we apply the key element of the Schwarzschild geometry, an absolute
rest system. Then the frequency shift factorizes in a natural way into
gravitational and (inverse!) time-dilation factors. As underlined elsewhere
\cite{APPB} this property is preserved in metric (\ref{dwie}).

Linet and Teyssandier \cite{LT} discussed the problem of frequency shift for a
broader class of potentials, namely axisymmetric potentials for rotating
gravitational field source. The weak-field and slow-motion case was considered
and the expansion for the Doppler shift was given up to the terms of an order
$1/c^4$.
The frequency shift discussed in \cite{LT} is:
\begin{equation}
{\omega_B\over\omega_A}={U^t_B\over U^t_A}{(1+u^i_Bl_i(B))\over(1+u^i_Al_i(A))},
\label{gg}
\end{equation}
where the normalized time component of the wave vector is conserved,
$l_t(B/A)=1$, (cf. (\ref{kat})), and the product of spatial components is
expressed in terms of coordinate velocity components, $u^i=U^i/U^t$,
\begin{equation}
  {U^il_i\over U^tl_t}=u^il_i.
\label{ggg}
\end{equation}
In fact, expression (\ref{gg}) takes the form of the frequency shift
(\ref{e16}), in the case of the Schwarzschild metric, where factorization
(\ref{e11}) of $U^t$ and parameterization (\ref{e13}), via velocity components
(instead of coordinate velocity components) $v^i$ (\ref{e12a},\ref{e12c}) is
used. In order to compare the Doppler shift expansions in reference \cite{LT}
and in this paper we expand $U^t$ (Eq. (\ref{e11})) up to the order $1/c^4$:
\begin{eqnarray}
  {1\over U^t}&=&\sqrt{1-{2GM\over c^2r}}\sqrt{1-{v^2\over c^2}}\approx
\nonumber
\\
  &&1-{1\over c^2}\left({GM\over r}+{1\over 2}v^2\right)
  +{1\over c^4}\left(-{1\over 2}\left({GM\over r}\right)^2
  +{1\over 2}{GM\over r}v^2-{1\over 8}v^4\right),
\end{eqnarray}
where the usual notation has been used, to find coincidence with corresponding
formula (76) in \cite{LT} (applied to the case of a Schwarzschild metric).
However, in the final expression (see Eq.(77) in \cite{LT}) the squared velocity,
$v^2$ (as measured by S), is substituted in \cite{LT} by the squared
coordinate velocity $u^2$. This leads to a difference in the Doppler shifts
expansions, one given in reference \cite{LT} and the other following our
approach, in the fourth order, $1/c^4$, as the relation between coordinate
velocity $u^i$ and velocity $v^i$ is $u^iu_i=g_{tt}v^iv_i$, and their
difference is of the order of $1/c^2$. It should be noted that expression
(\ref{e11}) is exact in this case (see also \cite{APPB}); it should also be
emphasized that the expansion of $U^t$ reveals inverse time-dilation
dependence, which has not been pointed out before (see, e.g.,
\cite{Bahder,GPS1,GPS2}).

The problem of the application and implementation of frequency shift
(\ref{e16}), in the framework of GPS, will be the subject of our following
paper; additionally the question of the Doppler shift expansion as given in
terms of velocity $v^i$ or in terms of coordinate velocity $u^i$ will be
discussed in detail.

\end{document}